\documentclass[preprint2]{aastex}

\def\tighttable{\def\baselinestretch{1.1}}
\def\arcsec{''}
\def\asec{\ifmmode ''\!. \else $''\!.$\fi}
\def\arcsecpoint{\ifmmode ''\!. \else $''\!.$\fi}
\def\micron{\ifmmode \mu{\rm m} \else $\mu$m\fi}

\def\Msun{M$_{\odot}$}

\def\HST{{\it HST}}
\def\sst{{\it Spitzer}} \def\I{{F814W}}
\def\Y{{F105W}}

\def\J{{F125W}}
\def\JH{{F140W}}
\def\H{{F160W}}
\def\cl{Abell 2744}
\def\zph{z_{\rm phot}}

\shorttitle{LBGs in Abell 2744}
\shortauthors{Zheng et al.}
\articleid{MS 94602}{} 

\begin{document}

\title{Young Galaxy Candidates in the Hubble Frontier Fields.\\I. Abell 2744}

\author{
Wei Zheng\altaffilmark{1},
Xinwen Shu\altaffilmark{2,3},
John Moustakas\altaffilmark{4},
Adi Zitrin\altaffilmark{5,6},
Holland C. Ford\altaffilmark{1},
Xingxing Huang\altaffilmark{1,3},
Tom Broadhurst\altaffilmark{7,8},
Alberto Molino\altaffilmark{9},
Jose M. Diego\altaffilmark{10},
Leopoldo Infante\altaffilmark{11},
Franz E. Bauer\altaffilmark{11,12,13},
Daniel D. Kelson\altaffilmark{14},
and
Renske Smit\altaffilmark{15}
}
\altaffiltext{1}{Department of Physics and Astronomy, Johns Hopkins
University, Baltimore, MD 21218}
\altaffiltext{2}{CEA Saclay, DSM/Irfu/Service d'Astrophysique, Orme des Merisiers, 91191 Gif-sur-Yvette Cedex, France}
\altaffiltext{3}{Department of Astronomy, University of Science and Technology 
of China, Hefei, Anhui 230026, China}
\altaffiltext{4}{Department of Physics and Astronomy, Siena College, Loudonville, NY 12211}
\altaffiltext{5}{Cahill Center for Astronomy and Astrophysics, California Institute of Technology, MS 249-17, Pasadena, CA 91125}
\altaffiltext{6}{Hubble Fellow}
\altaffiltext{7}{Department of Theoretical Physics, University of Basque Country UPV/EHU, Bilbao, Spain}
\altaffiltext{8}{IKERBASQUE, Basque Foundation for Science, Bilbao, Spain}
\altaffiltext{9}{Instituto de Astrof\'isica de Andaluc\'ia - CSIC, Glorieta de la Astronom\'ia, s/n. E-18008, Granada, Spain}
\altaffiltext{10}{IFCA, Instituto de F\'isica de Cantabria, UC-CSIC, s/n. E-39005 Santander, Spain} 
\altaffiltext{11}{Instituto de Astrof\'{\i}sica, Facultad de F\'{i}sica, Pontificia Universidad Cat\'{o}lica de Chile, 306, Santiago 22, Chile}
\altaffiltext{12}{Millennium Institute of Astrophysics, Vicu\~{n}a Mackenna 4860, 7820436 Macul, Santiago, Chile}
\altaffiltext{13}{Space Science Institute, Boulder, CO 80301}
\altaffiltext{14}{The Observatories of the Carnegie Institution for Science, Pasadena, CA 91101}
\altaffiltext{15}{Leiden Observatory, Leiden University, NL-2300 RA Leiden, The Netherlands}
\setcounter{footnote}{15}

\begin{abstract}
We report the discovery of 24 Lyman-break candidates at $7\lesssim z \lesssim 10.5 $, 
in the Hubble Frontier Fields (HFF)
imaging data of Abell 2744 ($z=0.308$), plus \emph{Spiter}/IRAC data
and archival ACS data.  
The sample includes a triple
image system with a photometric redshift of $z\simeq 7.4$. This high
redshift is geometrically confirmed by our lens model corresponding to
deflection angles that are 12\% larger than the lower-redshift systems
used to calibrate the lens model at $z= 2.019$.  The majority of our
high-redshift candidates are not expected to be multiply lensed given
their locations in the image plane and the brightness of foreground galaxies, 
but are magnified by factors of
$\sim 1.3 - 15$, so that we are seeing further down the luminosity function
than comparable deep field imaging. It is apparent that the redshift
distribution of these sources does not smoothly extend over the full
redshift range accessible at $z<12$, but appears to break above
$z=9$. 
Nine candidates are clustered within a small region of
$20\arcsec$ across, representing a potentially unprecedented concentration.
Given the poor statistics, however, we must await similar constraints from the 
additional HFF clusters to properly examine this trend.
The physical
properties of our candidates are examined using the range of lens
models developed for the HFF program by various groups including our
own, for a better estimate of underlying systematics.  Our
spectral-energy-distribution fits for the brightest objects suggest
stellar masses of $\simeq 10^{9}$~\Msun, star-formation rates of $\simeq
4$~\Msun~yr$^{-1}$, and a typical formation redshift of $z\lesssim
19$.  
\end{abstract}


\section{INTRODUCTION}

Our understanding of the first few billion years of cosmic time has
increased significantly in recent years, thanks to the {\em Hubble
  Space Telescope's} (\HST) Wide-Field Camera~3/Infrared Channel
\citep[WFC3/IR,][]{wfc3} as well as the {\em Spitzer Space
  Telescope's} Infrared Array Camera \citep[IRAC,][]{irac}. Until
recently, the Hubble Ultra Deep Field \citep{beckwith,garth} has provided our deepest view of the Universe, revealing a
considerable number of galaxy candidates at $z > 7$, including one
candidate at $z\simeq 10$ \citep{bouwens, bouwens1, ellis, oesch,
  garth}.  The cosmic epoch of $z\simeq 10$ is important to study as it
marks the dawn of galaxy formation and the beginning of reionization
of the intergalactic medium.  However, galaxies at that redshift are
extremely faint, making it difficult to discover and study the
abundant population of galaxies below $L^{*}$, the knee of the
luminosity function.  Fortunately, the magnification boost afforded by
gravitational lensing combined with \HST's exquisite imaging
capabilities in the near-infrared (NIR), provides an avenue for both
discovering and characterizing the intrinsic properties of galaxies
around $z\simeq 12$, when the Universe was less than half a billion years 
old. 

The Cluster Lensing And Supernova survey with Hubble
\citep[CLASH,][]{postman} carried out \HST{} imaging of 25 galaxy
clusters in 16 broad bands between $0.2-1.7$ $\mu$m to a depth of AB
magnitude $\sim 27$ with a total of 20 orbits per cluster.  The CLASH
program has led to many interesting discoveries of magnified,
intrinsically faint galaxies. Several hundred dropout galaxies have
been uncovered in the range $z\simeq 6-8$, with a few notable examples at
higher redshifts of $z\simeq 9-11$ \citep[see][]{zheng, bouwens2, coe,
  bradley}, helping to motivate dedicated deeper lensing surveys.  

The Hubble Frontier Fields (HFF) is a new initiative now being carried
out to observe the distant Universe to an unprecedented depth,
combining the power of deep \HST\ imaging and gravitational
lensing. In \HST's Cycles 21 and 22, 560 orbits of Director's
Discretionary Time have been allocated to observe four clusters. The
observations are carried out with four WFC3/IR filters (F160W, F140W,
F125W, F105W) and three ACS filters \cite[Advanced Camera for
  Surveys,][F814W, F606W, F435W]{acs}.  It is anticipated that 280
orbits will be allocated in Cycle 23 to observe two additional
clusters. In addition, deep \emph{Spitzer} and \emph{Chandra}
observations are planned for the six HFF fields.  These coordinated
observations will enable us to probe the star formation rate density
at $z\gtrsim 9$, study the faint end of the galaxy population at $z
\simeq 3-8 $, and map the dark matter in these clusters in
unprecedented detail via many multiple images of background sources
(Hubble Deep Fields Initiative 2012 Science Working Group
Report).\footnote{http://www.stsci.edu/hst/campaigns/frontier-fields/HDFI\_SWGReport2012.pdf}

We report the discovery of 24 candidate Lyman-break galaxies
(LBGs) at $z\gtrsim 7$ in the field of \cl, based on the HFF observations and 
archival data. The faintest sources detected are around AB magnitude 29.  
These objects, listed in Tables~\ref{tbl-z9}-\ref{tbl-z7}, have
``secure'' photometric redshifts greater than 7 and a negligible
probability ($<1\%$) of being at lower redshift.

We adopt a concordance cosmology with $\Omega_M=0.3$,
$\Omega_\Lambda=0.7$ and $h=H_0/100\,{\rm km\,s^{-1}\,Mpc^{-1}}=0.7$,
and the AB magnitude system throughout.

\section{DATA}

Abell 2744 ($z=0.308$) is the first HFF target in \HST's Cycle 21. It
is one of the most actively merging galaxy clusters known
\citep{merten}, displaying a large critical curve of roughly $
60\arcsec \times 30\arcsec$.  The six HFF clusters have been selected
to maximize the lensing boost, which means that systems with highly
complex mass distributions ({\it e.g.,} clusters in the process of
merging) have been selected \citep{torri,redlich,zitrin13}.  The
HFF observations of \cl\ (GO/DD 13495, PI: Lotz) were carried out
between 2013 Oct.~25 and 2014 Jul. 1. Additional WFC3/IR images obtained
in 2013 Aug. and 2014 Jun.-Jul. (GO~13386, PI:
Rodney) and ACS images of 2009 (GO 11689, PI: Dupke)
are retrieved from the Mikulski Archive for Space Telescopes
(MAST\footnote{\url{http://archive.stsci.edu/hst}}) and used.  Table~1 lists
the exposure times and limiting magnitudes for all the imaging used
in our analysis. 

\begin{figure}[h]\plotone{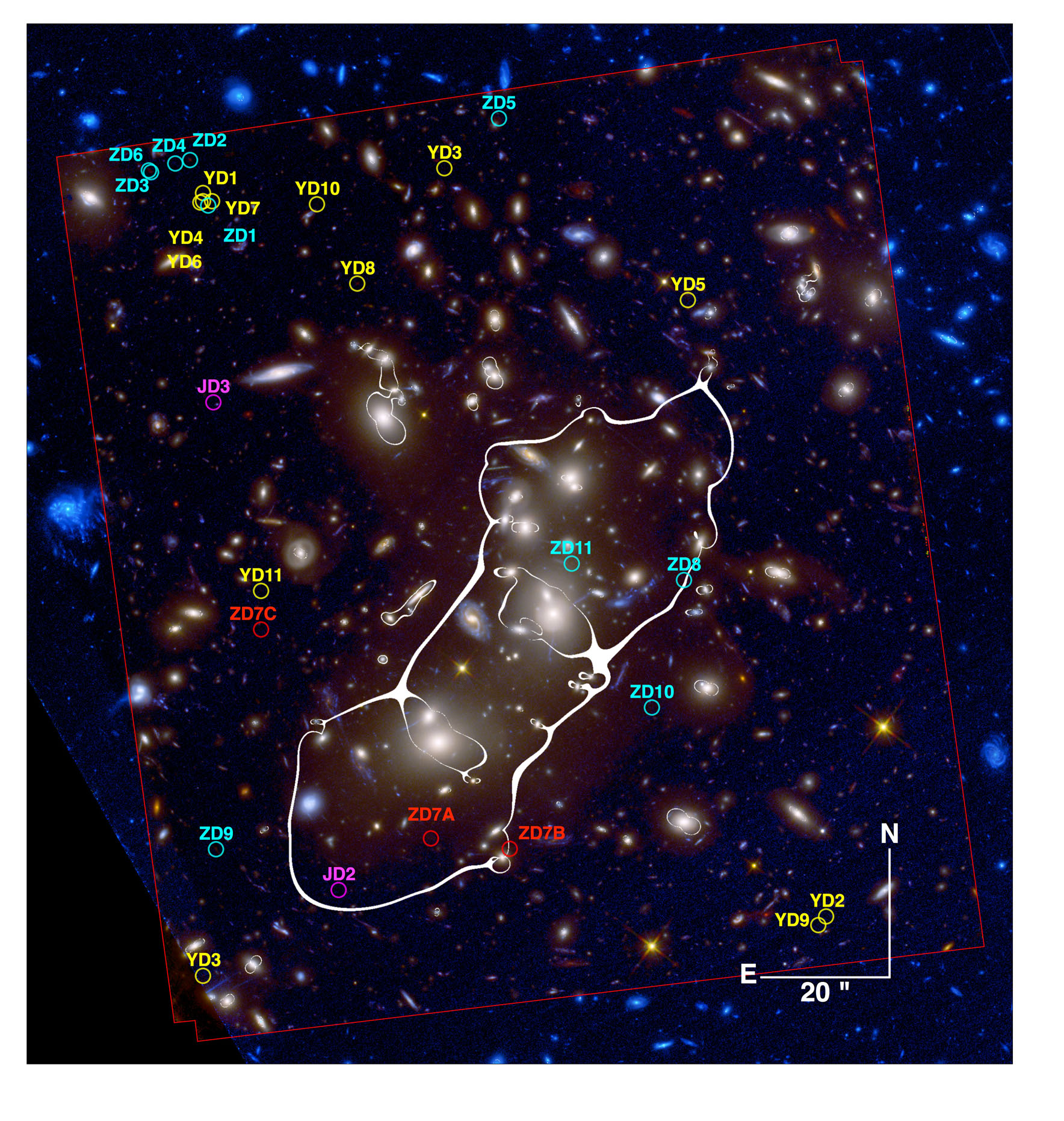}
\caption{Composite color image of \cl, made from the optical ACS images and 
the WFC3/IR F140W image.
The critical curves are from {\sc Zitrin ``NFW''} model (see \S 4.1) for 
background sources at $z = 9$ and
are plotted in white, marking the region with extreme magnification $\mu > 100$. 
The field of view of WFC3/IR is marked by a red box. 
Cyan symbols: $7 < z < 8$ objects; yellow: $8 \lesssim z < 9$, 
and magenta: $z>9$.
At the upper-left corner there is a region with four candidates at $z\simeq 
7.5$ and five at $z\simeq 8.5$. 
In the lower-left part a triple system of $z \simeq 7.4$ is marked in red.
}\label{fig-fov}
\end{figure}

We process the \HST\ data using {\tt APLUS} \citep{aplus}, an
automated pipeline modified from the {\tt APSIS} package
\citep{blakeslee} with an enhanced capability of processing WFC3 data
and aligning them with the ACS data. We retrieve the calibrated images
from the \HST\ instrument pipelines, namely the {\it flc} images for
ACS (corrected for the detector charge transfer efficiency) and {\it
  flt} images for WFC3/IR.  Recently, we have updated {\tt APLUS} so
that images of individual exposures are aligned using
\textit{DrizzlePac} \citep{astrodrz}, achieving an astrometric
precision of $\sim 0 \arcsecpoint 015$ or better. Figure~\ref{fig-fov} displays a 
composite color image of the cluster field.

Using {\tt APLUS}, we align, resample, and combine all the available
imaging in each filter to a common pixel scale of $0\arcsecpoint065$,
which is half of WFC3/IR's pixel scale and slightly larger than that of
ACS.  We then create detection images from the inverse-variance
weighted sum of the WFC3/IR and ACS images, respectively, and run {\tt
  SExtractor} \citep{bertin} in dual-image mode. The $z\gtrsim 7$ 
candidates are first selected from the NIR catalog derived from the WFC3/IR 
detection image, with a threshold of 1.5 times the signal-to-noise value over
a minimum of 4 pixels.  We choose colors
measured from isophotal magnitudes to select our high-redshift
candidates (see \S\ref{sec:selection}), as they balance the need
between depth and photometric precision \citep{ferguson}.  The
$5\sigma$ limiting magnitude in the WFC3/IR bands is $\sim 28.8$ in
a $0\arcsecpoint 4$ diameter aperture (see Table~1), and $\sim 29.2$
for the observed-frame optical ACS bands.

As part of the HFF campaign, deep \emph{Spitzer}/IRAC images were
obtained in 2013 Sep. and 2014 Jan.$-$Feb. in Channels 1 and 2 at wavelengths $3.1-3.9$ and
$3.9-5.0$ \micron, respectively, using Director's Discretionary Time
(Program 90257, PI: Soifer).  The effective exposure time in each channel, including that of 
the archival data (Program 84; PI: Rieke) obtained in 2004, is $\sim 339$~ksec. The IRAC corrected Basic Calibrated Data (cBCD) images are
processed with {\tt MOPEX} \citep{mopex} and sampled to a final pixel
scale of $0\arcsecpoint 6$.
In order to perform background matching of the individual cBCD frames we run all steps 
in the Overlap module, where we use the {\tt SExtractor} background estimation with a window 
size of 25 pixels. To create the mosaic images we run all three outlier modules ({\it i.e.}
Dual outlier, Mosaic Outlier and Box Outlier) and we use the default interpolation, with 
the fine resolution parameter = 0. The estimated $1 \sigma$ limiting
magnitude is $27.3$ for IRAC channel 1 (IRAC1, 3.6\micron) and $27.1$
for channel 2 (IRAC2, 4.5\micron; see Table~1).

\section{SELECTION}\label{sec:selection}

We search for LBGs using their distinct color around $0.1216 (1+z)$
\micron.  For example, at $z\simeq 7 - 8$, the Lyman break is at $\sim 1$ \micron, 
between the F814W  and F125W bands. 
Our selection criteria, in units of magnitude, are as follows: \newline
$\I - \Y > 0.8$\newline
$\Y - \J < 0.6$\newline
$\I - \Y > 0.8 + (\Y - \J)$\newline
These color cuts are similar to those utilized in previous work such as \cite{oesch10}.

For $z\simeq 8 - 9$, the break is at $\sim 1.15$ \micron, 
between the F105W and F140W bands: \newline
$\Y - \JH > 0.8$\newline
$\JH - \H < 0.6$\newline
$\Y - \JH > 0.8 + (\JH - \H)$

And for $z\simeq 10$, the break is between the F125W  and F160W bands: 
\newline $\J - \H > 0.8$

We require that a candidate
must not be detected above $1 \sigma$ in a summed image blueward of
the selection bands defined above. For objects at $z\simeq 7$, this
requires a non-detection in a summed image of the F606W and F435W
bands, while for candidates at $z\gtrsim 8$ this requires a
non-detection in the stacked optical detection image.  

In addition to the color selection criteria described above, we also
exclude candidates lying within one arcsecond of the detector edges,
in order to mitigate potentially spurious photometry.  We also exclude
candidates lying near stellar diffraction spikes, which are difficult
to remove because HFF WFC3/IR exposures were obtained at the same position angle. Finally, we also identify and remove
candidates with a color decrement of F160W - IRAC $> 3$, as they are most 
likely extremely red objects at lower redshift ($z\simeq 2$).

Our \HST{} photometry is measured within an isophotal aperture, but
aperture-corrected to a total flux using the $mag\_{auto} -
mag\_{iso}$ difference in the F160W band.  In a few cases where source
blending is significant, we visually inspect the images and choose an
aperture that is larger than the source's full-width at half-maximum
(FWHM), but not so large as to be affected by nearby sources, and use
the corresponding aperture magnitude in place of $mag\_{auto}$.  We
also verify that our aperture colors (and therefore our list of
high-redshift candidates) are not affected by image artifacts if we use 
the publicly
released \HST{} mosaics based on the \textit{Mosaicdrizzle} pipeline
\citep{koekemoer}.\footnote{http://archive.stsci.edu/pub/hlsp/frontier/abell2744/images/hst/v1.0}

The IRAC images of our candidates suffer from crowding due to the
instrument's large point spread function (PSF, FWHM
$\simeq 1\arcsecpoint 6$), such that simple aperture photometry might
result in inaccurate fluxes due to contamination from nearby
sources. To address this issue, we use a deblending technique whereby
contaminating neighbors are subtracted using {\tt GALFIT}
\citep{galfit} by performing a fit to the objects of interest and all
their close neighbors simultaneously in a
$\sim10\arcsec\times10\arcsec$ fitting window around the source 
of interest
\citep{overzier,zheng}. 
 The IRAC PSF is determined from the same
image by stacking 10 bright, isolated point sources. Positions and
radial profiles of neighboring sources in this region are derived from
the higher resolution \HST\ F160W-band mosaic, while the initial input
magnitudes are obtained by running {\tt SExtractor} on the IRAC
images. During the fitting process, all input parameters are allowed
to vary, except for the positions of the objects of interest.  
This process is similar to that of \cite{labbe06} and \cite{labbe10}. Four of
our candidates are so heavily blended by nearby bright sources with
complex radial profiles that {\tt GALFIT} fails to satisfactorily
deblend them. This yields a total of 16 sources for which photometry
or upper limits from {\tt GALFIT} are possible (see Table~\ref{tbl-irac}). 

\begin{figure}[h] \plotone{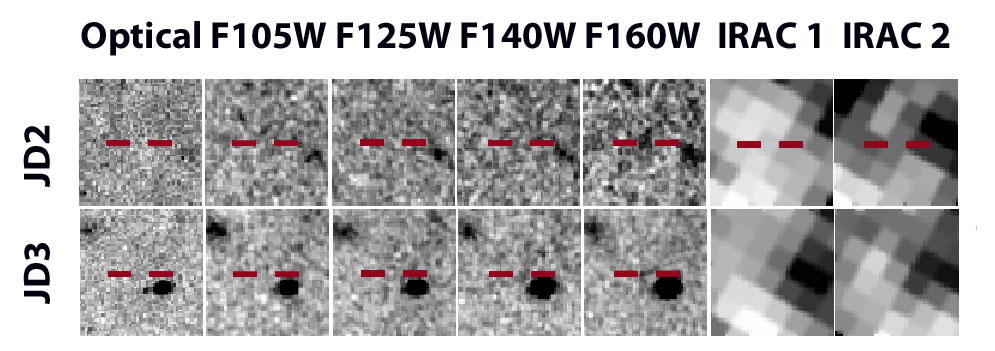}\caption{Cutout images of LBG candidates 
of $z > 9$ in \cl. 
The optical images are from the respective ACS detection images, which are
the weighted sums of ACS data in the F814W, F606W and F435W bands. 
Each candidate is at the image center, marked by pairs of red bars.
 For JD3, no red bars are present as no photometry is made because of a bright 
nearby source. 
Using the \HST\ photometry of this bright source, we estimate the
fluxes at 3.6 and 4.5 \micron\ with SED models and find that the best-fit model may account
for all the observed IRAC fluxes. Therefore it is likely that source JD3 is weak in
the two IRAC bands 
The field of view is 3\arcsecpoint 3, north is up and east to the left.
}\label{fig-z9}
\end{figure}

We carry out extensive tests to check the reliability of our IRAC
photometry for each source.  First, we place a simulated point source
of magnitude 25 near the candidate and run {\tt GALFIT} with different
fitting windows and background levels until the expected magnitude of
each simulated source is recovered (with a magnitude difference
$<0.1$~mag compared to the input value, consistent with the
photometric errors). We then proceed to fit the flux of each
candidate, using the fitting window and background level on the image
that recovered the brightness of the simulated source. We repeat these
tests at three different positions for the simulated source to verify
our measurement of the source magnitude.  To account for the
uncertainties in estimating the background, we choose the average magnitude 
of three measurements as the source magnitude,
which are reported in Table~\ref{tbl-irac}.  

\begin{figure}[h] \plotone{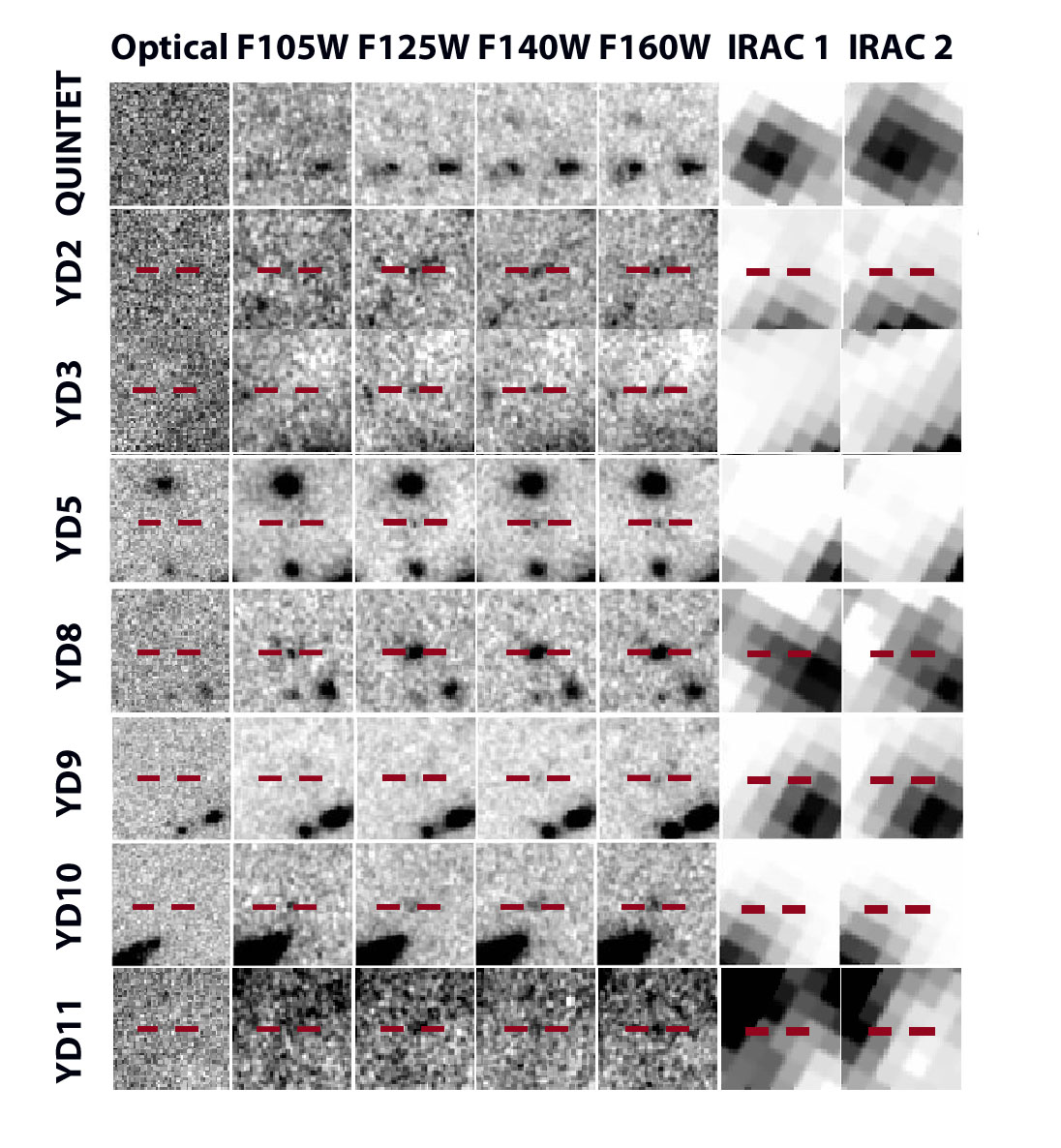}
\figcaption{Cutout images of LBG candidates 
of $8 \lesssim z < 9$ in \cl. 
The symbols are the same as Figure 2.
For the ``Quintet'' field, there are multiple candidates: YD1, 
YD4, YD6, YD7 and ZD1 (see Figure~7 for identification). 
For other fields, each candidate is at the image center, marked by pairs of 
red bars. For YD3 and YD5, no red bars are present as no photometry is made. 
The field of view is 3\arcsecpoint 3, north is up and east to the left.
}\label{fig-z8}
\end{figure}

\begin{figure}[h] \plotone{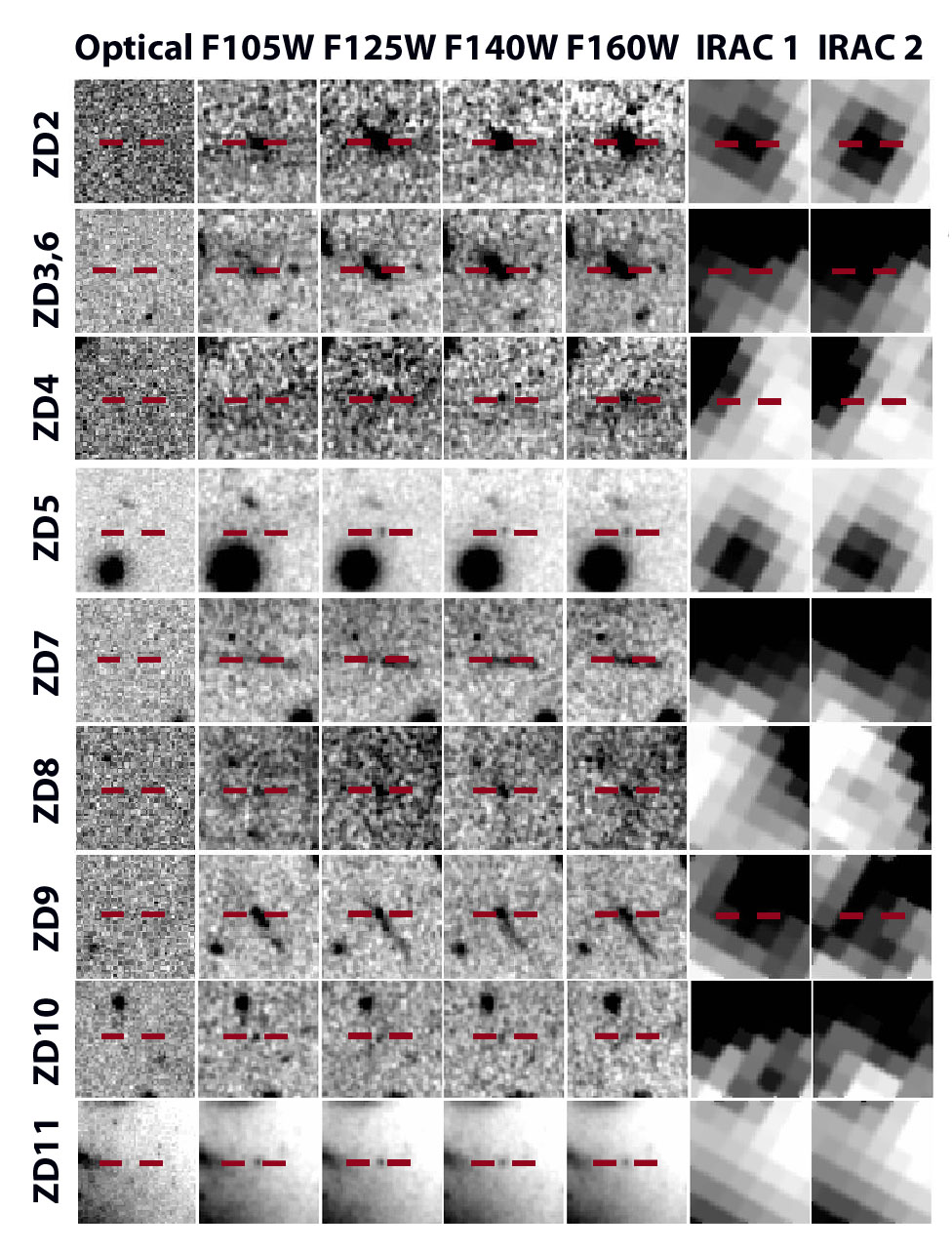}\caption{Cutout images of LBG candidates 
of $7 \lesssim z < 8$ in \cl. The symbols are the same as Figure 3.
}\label{fig-z7}
\end{figure}

In Figures~\ref{fig-z9}-\ref{fig-z7} we show cutout images of all the candidates.

\section{MODELS}

\subsection{Gravitational Lensing Models}\label{sec:models}
As part of the HFF initiative, seven independently derived
gravitational lensing models of the Abell~2744 field were developed
and publicly released through the MAST archive.\footnote{For details,
please see the Acknowledgements and
\url{http://archive.stsci.edu/prepds/frontier/lensmodels}.} For our
analysis we adopt the {\sc Zitrin ``NFW''} model as our fiducial lensing
model, and utilize the other six models to help quantify the
systematic uncertainties in our magnification estimates. The {\sc
Zitrin ``NFW''} model assumes a pseudo-isothermal elliptical mass
distribution for each cluster galaxy, scaled by the galaxy luminosity,
and an elliptical NFW \citep{nfw} halo for the dark matter. In the case of \cl, 
two such elliptical NFW halos are used, centered on the two
central, brightest cluster galaxies to represent the
global dark-matter component. These are combined with the galaxies component to 
generate, via a long Monte-Carlo-Markov-Chain minimization, the best-fitting model
for the total projected mass (see \citealt{elgordo, zitrin13}, and
references therein).

To estimate the systematic uncertainty in the magnification of each of
our high-redshift candidates, we exclude the highest and lowest
magnification factors and then calculate the difference in the
second-highest and second-lowest magnifications from the different models.
This procedure is  designed to mitigate potential extremes in the model predictions, and
better reflect the true systematic uncertainties.
The magnification
factors listed in Tables~\ref{tbl-z9}-\ref{tbl-z7} are the best-fit values at the
tabulated redshift based on the {\sc Zitrin ``NFW''} model, while the uncertainties are the quadrature sum of the
systematic and statistical uncertainties.  Figure~\ref{fig-fov} shows
the composite color image of the Abell~2744 field, overlaid with the
critical curves and identification numbers for all our candidates.

\subsection{Photometric Redshifts}\label{sec:photoz}

We calculate photometric redshifts using the Bayesian photometric
redshift code {\tt BPZ} \citep[Bayesian Photometric
Redshifts;][]{bpz,coe06}, adopting the same template library used by the
CLASH collaboration \citep{jouvel}. The template set consists of five
elliptical galaxy templates, two spiral galaxy templates, and four
starburst galaxy templates with moderately strong emission lines. The
templates were originally based on the P\'{E}GASE stellar population
synthesis models \citep{pegase}, but have been recalibrated using
spectroscopic redshifts of galaxies with deep, multiband photometry
from the FIREWORKS survey \citep{fireworks}.  We assume ignorant
({\it i.e.}, flat) priors on both galaxy type and redshift in the range
$z=0-12$.

Using BPZ, we identify 24 candidates that satisfy our color selection
criteria (see \S\ref{sec:selection}) and whose photometric redshifts
place them at $z>7$.  In Tables~\ref{tbl-z9}-\ref{tbl-z7} we list the 
coordinates, photometric redshifts, \emph{HST} photometry, and magnifications of
these 27 candidates. 

In order to infer the physical properties of our high-redshift
candidates (see \S\ref{sec:properties}), and as an additional check on
the BPZ-based photometric redshifts, we use the Bayesian spectral
energy distribution (SED) modeling code {\tt iSEDfit}
\citep{ised}. Using a Monte Carlo technique, we generate $20,000$
model SEDs with a broad range of star formation histories, ages,
stellar metallicities, dust content, and nebular emission-line
strength.  We use the Flexible Stellar Population Synthesis models
\citep[FSPS, v~2.4;][]{conroy09, conroy10} based on the {\sc miles}
stellar library \citep{sanchez-blazquez06} and assume the
\citet{chabrier} initial mass function from $0.1-100~M_{\odot}$.  We
adopt \emph{delayed} star-formation histories, ${\rm SFR}(t)\propto t
e^{-t/\tau}$, where SFR is the star formation rate, $t$ is the time
since the onset of star formation (``age''), and $\tau$ is the
characteristic time for star formation.  The advantage of this
parameterization is that it allows for both linearly rising
($t\ll\tau$) and exponentially declining ($t\gtrsim\tau$) star
formation histories, which may be important for modeling the SEDs of
galaxies at the highest redshifts \citep[e.g.,][]{papovich11}.  For
our photometric redshift calculations we adopt uniform priors on age
$t\in[0.01,12]$~Gyr,\footnote{Note that the age of the stellar
  population is never allowed to be older than the age of the Universe
  at the redshift under consideration.} star formation timescale
$\tau\in[0.01,5.0]$~Gyr, stellar metallicity
$Z/Z_{\odot}\in[0.04,1.6]$, and rest-frame $V$-band attenuation
$A_{V}\in[0-3]$~mag, assuming the time-dependent attenuation curve of
\cite{charlot00}.  Each model also includes nebular emission lines
whose luminosity is tied self-consistently to the number of
hydrogen-ionizing photons.

We find that {\tt iSEDfit} and BPZ yield statistically consistent
photometric redshifts for the majority of the candidates; the mean
difference is $\Delta z=0.08\pm0.24$ ({\tt iSEDfit} minus BPZ), which
is well within our quoted photometric redshift uncertainties.  In a
few cases {\tt iSEDfit} prefers a lower-redshift solution, $z\approx
2$; however, in every case these lower-redshift solutions require a
highly unlikely combination of physical properties, namely low stellar
masses, low star-formation rates, and large amounts of dust
attenuation.  Secondary peaks in the redshift probability distribution
from {\tt iSEDfit}, on the other hand, place these candidates at
$z>7$, in agreement with BPZ's primary redshift probability peaks.

\subsection{Multiple Systems}\label{sec:redarc}

To help corroborate the high-redshift nature of our 24 candidates, we
search for potential counter images near the locations predicted by
the gravitational lensing model.  Among the candidates that are inside or near the $z=7$
critical curves, JD1 \citep{zitrin14} and ZD7 are the only two cases where multiple images are found. 
For the others, the predicted counter images are either behind bright foreground galaxies, or too
faint to be confirmed.
Two of the counter images predicted for source ZD10 are near that of ZD11, which have been discussed 
by A14 (system 5).

\begin{figure}[h] \plotone{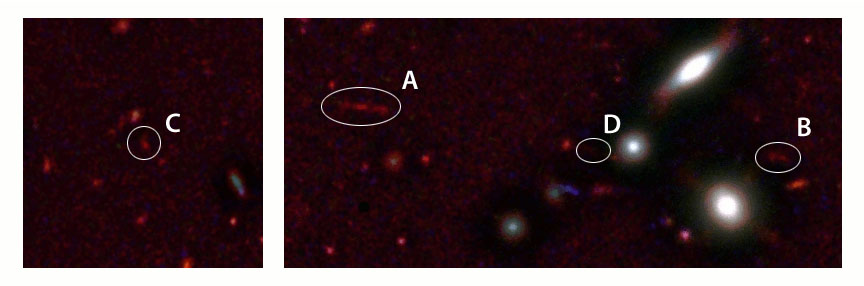} \caption{Red arclet ZD7 (marked as A) at redshift 7.4 and its two 
counter images B and C. Both images have been processed to boost the signal-to-noise of faint red objects 
and to reduce the contribution from nearby bright galaxies. An extremely faint source, 
marked as D, may be the potential fourth counter image.
}\label{fig-arc}
\end{figure}

The triple system ZD7 is shown in Figure~\ref{fig-arc}
and Table~\ref{tbl-ms}. Image A is an arclet made of two components that are
separated only by $\sim 1.5$ kpc in the source plane, assuming a
magnification of six.  This is similar to a case in Abell~1689 where a pair of LBGs at $z\simeq 7.6$ may be
merging \citep{a1689}. Image B is behind a bright foreground galaxy,
which we subtract before carrying out photometry. We
make use of the pure geometric scaling induced by strong lensing to
estimate a purely geometric distance for this triply imaged case. The
{\sc Zitrin ``NFW''} lensing model described in \S\ref{sec:models} is
based on eleven sets of multiply lensed galaxies between $2<z<4$,
including a spectroscopic redshift of system~6 at $z=2.019$ (J. Richard, 
in preparation\footnote{http://www.stsci.edu/hst/campaigns/frontier-fields/FF-Data}). This
spectroscopic redshift provides a normalization of the model so that
the deflection field induced in the lens plane,
$\vec{\alpha_L}(\vec{\theta})$, can be scaled to any redshift via the
lensing source distance ratio $f(z)=d_{ls}(z)/d_s(z)$ to provide the
observed deflection field
$\vec{\alpha}(\vec{\theta})={d_{ls}(z)/d_s(z)}\vec{\alpha_L}(\vec{\theta})$.
Hence only a simple scaling of the relative lensing distance ratios is
required to relate deflections at any given redshift to the lensing
distance of the normalization used to calibrate the lens model, which
in our case is $f(z)/f(z=2.019)$.  We find that this factor is $\sim
1.12$ for the triple system which minimizes the location of the
observed images relative to that generated by the model, and this
corresponds to a best BPZ estimate of $z\simeq 7.4$. A possible faint
fourth image D is noted between images A and B, close to another bright
galaxy (Figure~5).

\section{DISCUSSION}

\subsection{Individual Candidates}

Our paper serves as an independent verification of dropout objects in
other reports, including the recent work of \cite{atek} (A14
hereafter), \cite{laporte, coe14} and \cite{lam}.  Most of the 15 candidates in A14 are
at $6<z<7$, and therefore have only limited overlap with ours.  Three of
their candidates have been independently identified by us, but with
somewhat different photometric redshifts. Object ZD9 in Table~\ref{tbl-z7}, 
with $\zph=7.0$, corresponds to A14's object 561 ($z_{\rm A14}=7.5$).
Object ZD2 ($\zph=7.9$) is A14's object 2070 ($z_{\rm A14}=8.35$),
object Y1 ($z = 7.98$) in \cite{laporte} and object  2493-2561 in \cite{coe14}. 
Object ZD11 ($\zph=7.0$) is A14's object 5.2 ($z_{\rm A14}=6.4$) 
and system 17 ($\zph = 6.75$) in \cite{lam}.
The close pair of ZD3/ZD6 ($\zph = 7.7$) are marked by A14 as object 2070 
($z_{\rm A14}=8.35$) and object 2555-2516 in \cite{coe14}.
Objects YD4, YD7, YD8, YD10, ZD2, ZD3, ZD5 are objects 2493-2561, 2481-2561, 2306-3090, 2355-2566, 
2508-2497, 2555-2516, 2136-2432 in \cite{coe14}, respectively.

The IRAC2 flux of object ZD2 (Y1 in \citealt{laporte}) is more than
three times the IRAC1 value, which suggests a strong Balmer break.  A
similar case at $z\simeq 6$ was reported in Abell~383 \citep{a383} in
which both the IRAC1 and IRAC2 magnitudes are higher than the
\HST\ photometry by 1.6 magnitude.  At redshift $z\simeq 8$, the
Balmer break is in the IRAC1 band, and the model prediction is higher
than the measured IRAC1 flux.

\subsection{Physical Properties}\label{sec:properties}

In addition to reporting on the discovery of our high-redshift
candidates, we can also begin to characterize their physical
properties.  We defer a more detailed analysis of the full sample to a
forthcoming paper.
Here, we focus on the ten objects
at $z>7$ with the highest-confidence photometric redshifts which have
well-measured photometry in at least one IRAC channel.  Photometry
redward of the Balmer break is particularly important for placing
meaningful constraints on the stellar mass and age of the stellar
populations in these distant objects.

\begin{figure}[h] \plotone{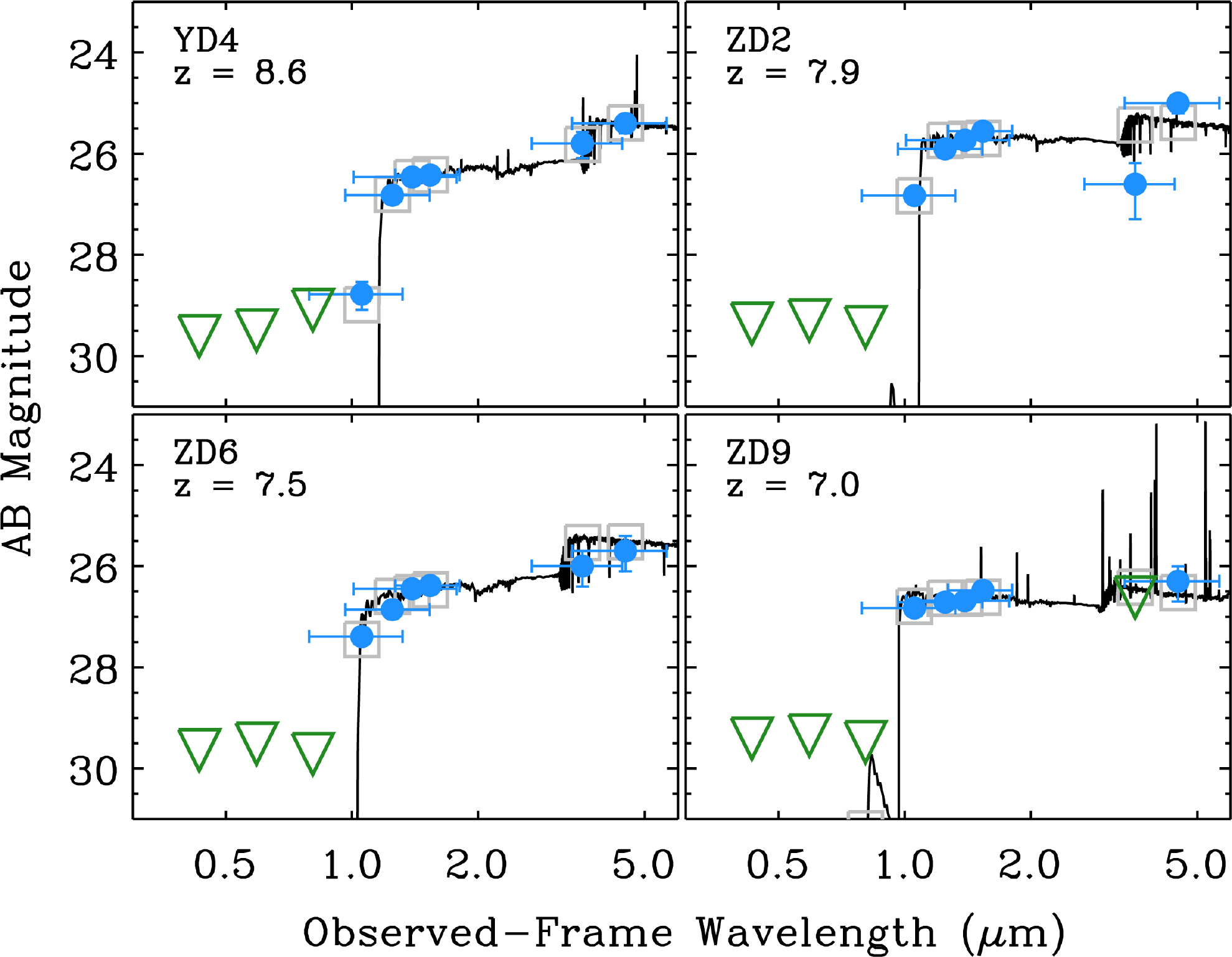}\caption{
Observed-frame SEDs of four bright candidates that have 
well-measured IRAC photometry. The filled blue points show the
observed photometry, while the open green triangles indicate $2\sigma$
upper limits.  The black spectrum shows the best-fitting (maximum
likelihood) SED based on our Bayesian SED modeling using {\tt
  iSEDfit}.  The large gray squares show the photometry of the
best-fitting model convolved with the ACS, WFC3, and IRAC filter
response curves.}
\label{fig-sed}
\end{figure}

To infer the physical properties of these galaxies we use {\tt
iSEDfit}, but adopt a more restricted set of priors than the ones
used to estimate photometric redshifts (see \S\ref{sec:photoz}).
Specifically, we adopt uniform priors on galaxy age
$t\in[10,750]$~Myr, $\tau\in[10,1000]$~Myr, stellar metallicity
$Z/Z_{\odot}\in[0.04,1.0]$, and we assume no dust attenuation
\citep[{\it e.g.},][]{bouwens3}.  Recall that the age of the Universe
at $z=7-8$ in our adopted concordance cosmology is $750-630$~Myr.

We find that our $z>7$ candidates have demagnified stellar masses of
around $10^{9}$~\Msun, and SFRs of approximately $4$~\Msun~yr$^{-1}$.
These results imply an average doubling time of around $500$~Myr,
which is comparable to the age of the Universe at
$z \simeq 8$.\footnote{The doubling time is the time it would take to 
double the stellar mass of the galaxy, where we have assumed that
$50\%$ of the stellar mass formed is returned to the interstellar
medium via supernovae and stellar winds.}  The ages of the galaxies
in our sample are less well constrained given the uncertainties in our
IRAC photometry; nevertheless, we find a median SFR-weighted age for
the sample of $\lesssim 430$~Myr ($95\%$ confidence interval),
corresponding to a typical formation redshift of $z\lesssim 19$.
Figure~\ref{fig-sed} presents the SEDs of four representative galaxies in our
sample, sorted by decreasing redshift, as well as the maximum
likelihood fits derived using {\tt iSEDfit}.

\subsection{Source Clustering}

An apparent concentration of candidates northeast of the cluster
center is shown in Figure~\ref{fig-fov}. Nine objects at $z \simeq 7-8$ are found
within a region of $20\arcsec$. Since the average magnification in
that area is not high ($\mu \simeq 1.4$), this apparent
overdensity is likely intrinsic rather than being due to lensing.  In
addition, Figure~7 shows a small region ``Quintet'' where five candidates
are located within approximately $2\arcsec$ of one another: objects
YD1, YD4, YD6, YD7 and ZD1. These objects have similar estimated
photometric redshifts of between 7.9 and 8.6, and their projected
separations in the source plane are within $\sim8$~kpc. Given the 
uncertainties in our
photometric redshifts, it is therefore possible that these sources are
physically associated.  

\begin{figure}[h] \plotone{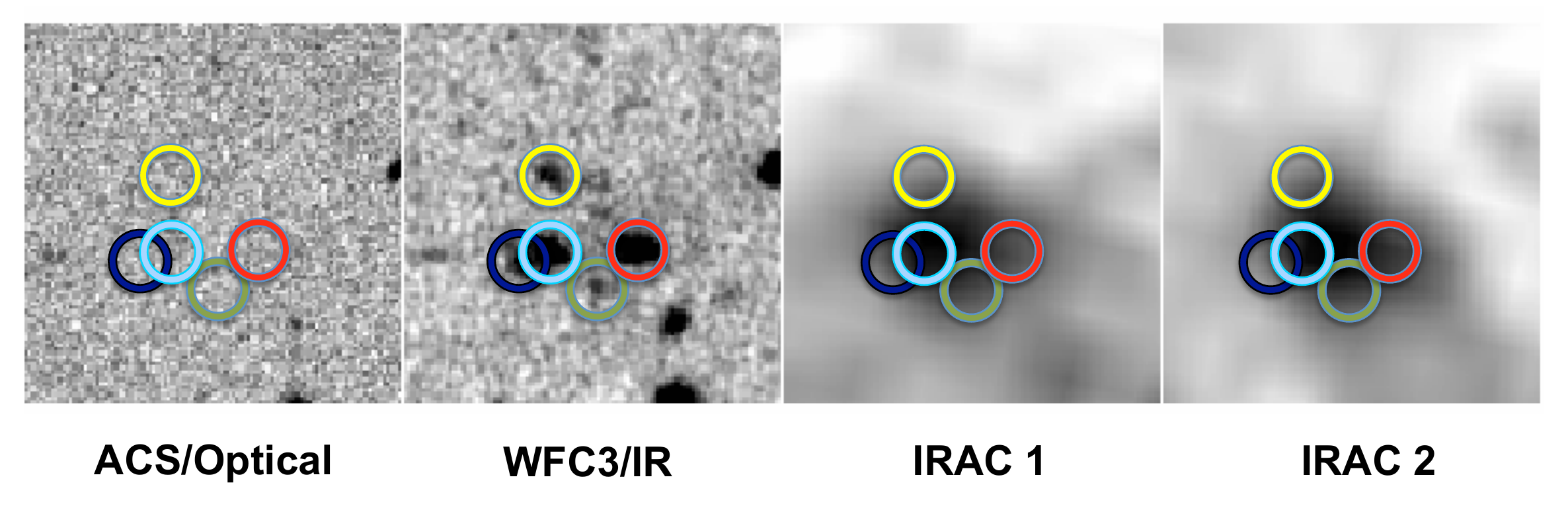} \caption{``Quintet'' of 
LBGs at $\zph \simeq 8.3 $, in an expanded view of the first row in Figure~2.  
The magnification factor is $\simeq 1.4$ and separations are $< 8$ kpc in the source plane. 
The images are $6\arcsecpoint 5$ on each side. The yellow circle marks object YD1; 
light blue: YD4; dark blue: object YD6; red: object YD7; 
and green: object ZD1. The circle size is approximately $0\arcsecpoint 9$ in diameter.
The two \HST\ images are the detection images summed over ACS and 
WFC3/IR bands, respectively. With an IRAC PSF source diameter of $1\arcsecpoint 6$ 
(approximately twice the circles'), these sources are considerably blended. 
We are able to derive their fluxes or upper limits, assuming
fixed source positions in {\tt GALFIT} fitting, 
}\label{fig-quintet}
\end{figure}

To estimate the uncertainties in IRAC photometry for these sources whose separations
are close to the IRAC PSF size, 
we place 500 sets of simulated sources of different brightness and compared the input and fitted fluxes.
In more than 68\% of the cases, the fitted results are within 0.5 mag of the 
inputs for three brighter sources in the F160W band, while 
for the faintest source the output flux decreases with the F160W-IRAC color. 
Due to the source confusion, our simulations suggest that the de-blending with 
{\tt GALFIT} works better for the brighter objects, while it is biased to fainter 
objects in the sense that their fluxes are likely to be overestimated. 
This is because that {\tt GALFIT} often crashes under a very faint flux 
level.
We therefore conclude that we are able to deblend these sources with reasonable accuracy,
except for the faint source ZD1 with large uncertainties.

Overdensities in the high-redshift domain have been previously
reported.  For example, \citet{trenti} identified four candidates with
$z\simeq 8$ within $70\arcsec$ in the Field~BoRG58.  \cite{a1703}
found seven LBGs at $z\simeq 7$ in the WFC3/IR field ($\sim 120\arcsec
\times 130\arcsec$) of Abell 1703.  However, our finding of nine LBGs
at $7<z<9$ within $20\arcsec$ ($\sim 80$~kpc in the source plane) is
unique, suggesting that the cosmic variance in source density
\citep{cv} is more significant than anticipated. 
It appears that cosmic variance increases with redshift \citep{moster, bouwens4}, as evidenced by different results from
surveys.
This trend might explain both the large number of clustered 
$z\simeq 8$ sources and the deficiency of $z\gtrsim 9$ sources and illustrate the need for observations in more fields.

\section{CONCLUSION}

We find 24 LBG candidates at $7 \lesssim z \lesssim 10.5$ in the HFF 
imaging of \cl, reaching an intrinsic magnitude of 
$\sim 32$. One source at $z\simeq 7.4$ is lensed into three images.
Significant clustering is observed on the intrinsic 
scale of $10-100$~kpc. Thanks to gravitational lensing,
we are able to carry out Spitzer/IRAC photometry for 16 of
the sources. SED fitting to the brightest candidates 
suggests stellar masses of $\simeq 10^9$~\Msun, star-formation 
rates of $\simeq 4$~\Msun\ per year, and a typical formation 
redshift of $z \simeq 19$.

The redshift distribution of our sample is not a smoothly declining function
towards higher redshift. In particular, our redshift distribution
does not extend smoothly beyond $z\simeq 9$, clustering notwithstanding,
Considering the effect of cosmic variance, the number density in our sample
is consistent with that derived from other studies, {\it e.g.}, \cite{bouwens4}. 
Given the level of clustering that we see in \cl, 
it will be  important to average over more HFF, and to perform a luminosity 
function analysis so that the redshift dependence can be better related to
galaxy mass.

The work presented in this paper is based on observations made with
the NASA/ESA {\it Hubble Space Telescope}, and has been supported by
award AR-13079 from the Space Telescope Science Institute (STScI),
which is operated by the Association of Universities for Research in
Astronomy, Inc. under NASA contract NAS~5-26555.  It is also based on
data obtained with the {\it Spitzer Space Telescope}, which is
operated by the Jet Propulsion Laboratory, California Institute of
Technology under a contract with NASA.  This work utilizes
gravitational lensing models produced by P.I.s Bradac, Ebeling, Merten
\& Zitrin, Sharon, and Williams, funded as part of the \HST\ Frontier
Fields program conducted by STScI.  These models were calibrated using
arcs identified in archival \HST{} imaging by \cite{merten},
spectroscopic redshifts of arcs obtained using the VLT/FORS2
spectrograph (J. Richard et~al. 2014, in prep.), and VLT and
Subaru/Suprimecam imaging of the Abell~2744 field \citep{cypriano,
  okabe08, okabe10, okabe10b}.  XS acknowledges support from
FP7-SPACE-2012-ASTRODEEP-312725, and NSFC grants 11103017 and
11233002.  Support for AZ is provided by NASA through Hubble
Fellowship grant HST-HF-51334.01-A awarded by STScI. FEB acknowledges
support from Basal-CATA PFB-06/2007, CONICYT-Chile grants (FONDECYT
1141218, ALMA-CONICYT 31100004, Gemini-CONICYT 32120003, Anillo
ACT1101), and the Millennium Institute of Astrophysics (Project IC120009).  
JMD acknowledges support from the Spanish consolider project 
CAD2010-00064 and AYA2012-39475-C02-01.
We thank M. Meneghitti for helpful comments. 

\hskip 1.5in

{
\begin{deluxetable}{ccccc}
\tablecaption{Summary of Observations\label{tbl-sum}}
\tablewidth{0pt}
\footnotesize
\tighttable
\tablehead{
\colhead{Telescope} &
\colhead{Band} &
\colhead{Date} &
\colhead{Exposure Time} &
\colhead{Limiting}
\\
\colhead{} &
\colhead{} &
\colhead{} &
\colhead{(sec)} &
\colhead{Magnitude ($5\sigma$)}
}
\startdata \HST  & F160W&2013 Aug.$-$2014 Jul.& 71664 & 28.7 \\ \HST  & F140W&2013 Oct.,Nov.& 28140 & 28.5 \\ 
\HST  & F125W&2013 Oct.$-$2014 Jul.& 36620 & 28.6 \\  \HST  & F105W&2013 Aug.$-$2014 Jul.& 71209 & 29.0 \\  \HST  & F814W&2014 May$-$Jul., 2009 Oct.& 117518 & 29.4 \\ \HST  & F606W&2014 May, 2009 Oct.& 36882 & 29.2 \\ \HST  & F435W&2014 Jun.$-$Jul., 2009 Oct.& 61909 & 29.3 \\ \sst  & IRAC1& 2013 Sep., 2014 Jan.,Feb.& 339291 & 25.5 \\
   &      & 2004 Jun.,Nov.& & \\ 
\sst  & IRAC2& 2013 Sep., 2014 Jan.,Feb.& 339291 & 25.3 \\ 
  &      & 2004 Jun.,Nov.& & \\ \enddata
\end{deluxetable}
}

{
\begin{deluxetable}{lccccccccc}
\tabletypesize{\tiny}
\tabletypesize{\scriptsize}
\setlength{\tabcolsep}{0.05in}
\rotate
\tablecaption{WFC3 and ACS Photometry\tablenotemark{a} of Candidates at $z > 9$
\label{tbl-z9}}
\tablewidth{0pt}
\tiny 
\tighttable
\tablehead{
\colhead{Name} &
\colhead{Photometric} &
\colhead{RA} &
\colhead{Dec} &
\colhead{F160W} &
\colhead{F140W} &
\colhead{F125W} &
\colhead{F105W} &
\colhead{F814W} &
\colhead{$\mu$\tablenotemark{b}} \\
\colhead{} &
\colhead{Redshift} &
\colhead{(J2000)} &
\colhead{(J2000)} &
\colhead{} &
\colhead{} &
\colhead{} &
\colhead{} &
\colhead{} &
\colhead{} 
}
\startdata
JD2\tablenotemark{c} & $10.5\pm 0.7$ &   3.597037 & $-30.412132$ & $ 28.00 \pm  0.15 $& $  29.00 \pm 
 0.43 $& $  30.20 \pm  1.33 $& $  29.52 \pm  0.52 $& $ > 31.00 $& $14.8 ^{+66} _{-7.8}$ \\  JD3 & $10.0^{+0.7}_{-1.0}$ &  3.603344 & $-30.390978$ & $  28.68 \pm  0.16 $& $ 29.26 \pm  0.30 $& $ 30.30 \pm 0.78 $& $ > 30.50 $& $ > 30.50 $& $1.6 ^{+0.9} _{-0.1}$   \enddata
\tablenotetext{a}{Magnitudes are isophotal,
scaled by an aperture correction term derived in the \H\ band. The errors and 
limiting magnitudes are $1\sigma$. Photometric redshifts have been derived using BPZ, and the quoted
uncertainties indicate the 68\% confidence interval.}
\tablenotetext{b}{Magnification factor from {\sc Zitrin ``NFW''} model (see \S \ref{sec:models}), 
where uncertainties are combined 
from two sources: (1) the model statistical uncertainties, which are relatively small; and more prominently, 
(2) the maximum differences with other six models, excluding one highest and one lowest values.}
\tablenotetext{c}{For system JD1, see \cite{zitrin14}}
\end{deluxetable}
}

{
\begin{deluxetable}{lccccccccc}
\tabletypesize{\tiny}
\tabletypesize{\scriptsize}
\setlength{\tabcolsep}{0.05in}
\rotate
\tablecaption{WFC3 and ACS Photometry\tablenotemark{a} of Candidates at $8 \lesssim z < 9$
\label{tbl-z8}}
\tablewidth{0pt}
\tiny 
\tighttable
\tablehead{
\colhead{Name} &
\colhead{Photometric} &
\colhead{RA} &
\colhead{Dec} &
\colhead{F160W} &
\colhead{F140W} &
\colhead{F125W} &
\colhead{F105W} &
\colhead{F814W} &
\colhead{$\mu$\tablenotemark{b}} \\
\colhead{} &
\colhead{Redshift} &
\colhead{(J2000)} &
\colhead{(J2000)} &
\colhead{} &
\colhead{} &
\colhead{} &
\colhead{} &
\colhead{} &
\colhead{} 
}
\startdata
YD1 & $8.7^{+0.5}_{-0.2}$ &   3.603856 & $-30.381905$ & $  27.83 \pm  0.08 $& $  27.86 \pm  0.09 $& $  28.49 \pm  0.17 $& $  29.85 \pm  0.43 $& $> 31.00 $& $1.4 ^{+0.7} _{-0.1}$ \\  YD2 & $8.3\pm 0.2$ &   3.572515 & $-30.413267$ & $  28.12 \pm  0.12 $& $  28.11 \pm 
 0.13 $& $  28.11 \pm  0.13 $& $  29.84 \pm  0.48 $& $ > 31.00 $& $1.6 ^{+0.4} _{-0.1}$ \\  YD3 & $8.8^{+0.4}_{-0.2}$  &  3.603858 & $-30.415842$ & $  28.33 \pm  0.10 $& $  28.63 \pm  0.15 $& $  28.66 \pm  
0.18 $& $ > 30.50 $& $ > 31.00 $& $2.9 ^{+0.5} _{-0.6}$ \\ YD4\tablenotemark{c} & $8.5 \pm 0.1$ &   3.603864 & $-30.382265$ & $  26.42 \pm  0.04 $& $  26.46 \pm  0.04 $& $  26.82 
\pm  0.06 $& $  28.78 \pm  0.27 $& $ > 31.00 $& $1.4 ^{+0.7} _{-0.1}$ \\  YD5 & $8.5\pm 0.3$ &   3.579479 & $-30.386534$ & $  27.83 \pm  0.10 $& $  27.66
 \pm  0.09 $& $  28.38 \pm  0.18 $& $  29.24 \pm  0.29 $& $  > 31.00 $& $3.0 ^{+9.8} _{-0.9}$ \\  YD6 & $8.3 \pm 0.2$ &   3.604005 & $-30.382309$ & $  26.95 \pm  0.06 $& $  27.09 \pm  0.07 $& $  27.36 \pm  0.10 $& $  28.83 \pm  0.26 $& $> 31.00 $& $1.4 ^{+0.7} _{-0.1}$ \\  YD7\tablenotemark{c} & $8.3 \pm 0.1$ &   3.603397 & $-30.382256$ & $  26.17 \pm  0.03 $& $  26.10 \pm  0.03 $& $  26.38 \pm  0.05 $& $  27.53 \pm  0.09 $& $ > 31.00 $& $1.4 ^{+0.7} _{-0.1}$ \\  YD8\tablenotemark{c} & $8.1\pm0.1$ &   3.596096 & $-30.385832$ & $  26.65 \pm  0.04 $& $  26.49 \pm  0.04 $& $  26.68 
\pm  0.04 $& $  27.76 \pm  0.09 $& $> 31.00 $& $1.9 ^{+5.6} _{-0.2}$ \\  YD9 & $8.3^{+0.2}_{-0.6}$  &  3.572902 & $-30.413658$ & $  28.49 \pm  0.14 $& $  28.22 \pm  0.12 $& $  28.68 \pm  
0.18 $& $  29.62 \pm  0.32 $& $ > 31.00 $& $1.6 ^{+0.5} _{-0.1}$ \\ YD10\tablenotemark{c} & $8.3\pm 0.2$ &   3.598108 & $-30.382393$ & $  27.70 \pm  0.09 $& $  27.57 \pm
  0.09 $& $  27.89 \pm  0.12 $& $  28.89 \pm  0.23 $& $> 31.00 $& $1.5 ^{+2.3} _{-0.1}$ \\  YD11 & $8.0^{+0.3}_{-0.6}$  &  3.600947 & $-30.399149$ & $ 28.96 \pm  0.17 $& $  28.97 \pm  0.19 $& $  28.89 \pm  
0.18 $& $ > 30.50 $& $ > 31.00 $& $3.4 ^{+0.9} _{-0.1}$  \\ \enddata
\tablenotetext{a}{Magnitudes are isophotal,
scaled by an aperture correction term derived in the \H\ band. The errors and 
limiting magnitudes are $1\sigma$. Photometric redshifts have been derived using BPZ, and the quoted
uncertainties indicate the 68\% confidence interval.}
\tablenotetext{b}{Magnification factor from {\sc Zitrin ``NFW''} model (see \S 4.1), 
where uncertainties are combined 
from two sources: (1) the model statistical uncertainties, which are relatively small; and more prominently, 
(2) the maximum differences with other six models, excluding one highest and one lowest values.}
\tablenotetext{c}{Also in \cite{coe14}.}
\end{deluxetable}
}

{
\begin{deluxetable}{lccccccccc}
\tabletypesize{\tiny}
\tabletypesize{\scriptsize}
\setlength{\tabcolsep}{0.05in}
\rotate
\tablecaption{WFC3 and ACS Photometry\tablenotemark{a} of Candidates at $7 \lesssim z < 8$
\label{tbl-z7}}
\tablewidth{0pt}
\tiny 
\tighttable
\tablehead{
\colhead{Name} &
\colhead{Photometric} &
\colhead{RA} &
\colhead{Dec} &
\colhead{F160W} &
\colhead{F140W} &
\colhead{F125W} &
\colhead{F105W} &
\colhead{F814W} &
\colhead{$\mu$\tablenotemark{b}} \\
\colhead{} &
\colhead{Redshift} &
\colhead{(J2000)} &
\colhead{(J2000)} &
\colhead{} &
\colhead{} &
\colhead{} &
\colhead{} &
\colhead{} &
\colhead{} 
}
\startdata
ZD1 &  $7.4^{+0.3}_{-0.6}$ &   3.603582 & $-30.382442$ & $  28.52 \pm  0.19 $& $  28.09 \pm  0.14 $&
 $  28.47 \pm  0.21 $& $ > 30.50 $& $> 31.00 $& $1.4 ^{+0.7} _{-0.1}$ \\  ZD2\tablenotemark{c,d} & $7.9 \pm 0.1$ &   3.604520 & $-30.380472$ & $ 25.56 \pm  0.03 $& $  25.73 \pm
0.03 $& $  25.91 \pm  0.04 $& $  26.83 \pm  0.07 $& $ > 30.00 $& $1.3 ^{+0.7} _{-0.1}$ \\ ZD3\tablenotemark{c,d} & $7.7 \pm 0.1$ &   3.606477 & $-30.380993$ & $  26.45 \pm  0.04 $& $  26.57 \pm 0.05 $& $  26.64 \pm  0.05 $& $  27.50 \pm  0.08 $& $> 31.00 $& $1.3 ^{+1.0} _{-0.1}$ \\ ZD4 & $ 7.8\pm 0.3$ & 3.605263 & $-30.380606$ & $  27.97 \pm  0.12 $& $  27.95 \pm  0.13 $& $  28.29 \pm  
0.17 $& $  29.05 \pm  0.26 $& $> 31.00 $& $1.3 ^{+0.7} _{-0.1}$ \\  ZD5\tablenotemark{d} & $7.6 ^{+0.1}_{-0.3} $ &   3.588985 & $-30.378662$ & $  27.55 \pm  0.05 $& $  27.77 \pm  0.07 $& $  27.52 \pm  0.06 $& $  28.29 \pm  0.09 $& $>30.00 $& $1.6 ^{+0.9} _{-0.1}$ \\ ZD6\tablenotemark{c}  &  $7.5\pm 0.1$ &   3.606575 & $-30.380928$ & $  26.38 \pm  0.04 $& $  26.45 \pm
  0.04 $& $  26.86 \pm  0.06 $& $  27.39 \pm  0.08 $& $> 31.00 $& $1.3 ^{+1.1} _{-0.1}$ \\  ZD7 & $7.3 \pm 0.2 $ &   3.592285 & $-30.409911$ & $ 26.98 \pm  0.06 $& $  26.99 \pm  0.07 $& $  26.92 \pm  0.06 $& $  27.39 \pm 0.07 $& $> 30.50 $& $5.9 ^{+7.8} _{-3.0}$ \\  ZD8 & $7.5 \pm 0.3$ &3.579668 & $-30.398678$ & $ 28.33 \pm  0.12 $& $  28.11 \pm  0.11 $& $  28.29 \pm 
0.14 $& $28.89 \pm 0.17 $& $ > 31.00 $& $14.0 ^{+38} _{-6.8}$  \\ ZD9\tablenotemark{c} &  $7.0 \pm 0.1$ &   3.603208 & $-30.410368$ & $  26.48 \pm  0.04 $& $  26.68 \pm  
0.06 $& $  26.71 \pm  0.06 $& $  26.83 \pm  0.05 $& $  > 31.00 $& $3.4 ^{+7.8} _{-0.8}$ \\ ZD10 & $7.0 \pm 0.3$ & 3.581282 & $-30.404207$ & $ 28.75 \pm  0.16 $& $ 28.61 \pm  0.16 $& $  28.78 \pm  
0.19 $& $ 28.98 \pm 0.17 $& $ > 31.00 $& $9.8 ^{+15} _{-4.4}$  \\ ZD11\tablenotemark{c} & $7.0 \pm 0.1$  &  3.585321 & $-30.397964$ & $ 27.49 \pm  0.04 $& $  27.45 \pm  0.04 $& $  27.33 \pm  
0.04 $& $ 27.55 \pm 0.03 $& $ > 31.00 $& $4.7 ^{+1.8} _{-3.1}$ \\   \enddata
\tablenotetext{a}{Magnitudes are isophotal,
scaled by an aperture correction term derived in the \H\ band. The errors and 
limiting magnitudes are $1\sigma$. Photometric redshifts are BPZ with $1\sigma$ error.}
\tablenotetext{b}{Magnification factor from {\sc Zitrin ``NFW''} model (see \S 4.1), 
where uncertainties are combined 
from two sources: (1) the model statistical uncertainties, which are relatively small; and more prominently, 
(2) the maximum differences with other six models, excluding one highest and one lowest values.}
\tablenotetext{c}{Reported by A14 and \cite{laporte}.}
\tablenotetext{d}{Also in \cite{coe14}.}
\end{deluxetable}
}

{
\begin{deluxetable}{lcc}
\tablecaption{IRAC photometry for Selected Candidates \label{tbl-irac}}
\tablewidth{0pt}
\tiny 
\tighttable
\tablehead{
\colhead{Name} &
\colhead{IRAC 1} &
\colhead{IRAC 2}
\\
\colhead{} &
\colhead{} &
\colhead{} 
}
\startdata
JD2 &  $> 27.3 $& $> 27.1 $ \\  
YD1 &  $> 27.3 $& $  26.2 \pm  0.5 $ \\  
YD2 &  $> 27.3 $& $> 27.1 $ \\ 
YD4\tablenotemark{a} &  $  25.8 \pm  0.3 $& $  25.4 \pm  0.2 $ \\  
YD6 &  $  25.5 \pm  0.2 $& $  25.2 \pm  0.2 $ \\  
YD7\tablenotemark{a} &  $  26.5 \pm  0.6 $& $  26.2 \pm  0.5 $ \\  
YD8 &  $ 26.5 \pm 0.5 $& $> 27.1 $ \\ 
YD9 &  $> 27.3 $& $> 27.1 $ \\  YD10 &  $> 27.3 $& $> 27.1 $ \\  YD11 &  $> 27.3 $& $> 27.1 $ \\  ZD1 &  $  25.7 \pm  1.3 $& $  26.1 \pm  1.4 $ \\  
ZD2 &  $  26.6 \pm  0.7 $& $  25.0 \pm  0.2 $ \\  ZD3\tablenotemark{a} &  $  26.5 \pm  0.6 $& $  26.1 \pm  0.5 $ \\  
ZD4 &  $> 27.3 $& $> 27.1 $\\  ZD6\tablenotemark{a} &  $  26.0 \pm  0.4 $& $  25.7 \pm  0.4 $ \\  
ZD9 &  $ 26.6 \pm 0.4 $ & $26.6 \pm 0.3 $ \\  
\enddata 
\tablenotetext{a}{Two respective close pairs within one IRAC pixel, each fitted as one component.  
The fluxes of individual components are partitioned by a ratio of their fluxes in the F160W band.}
\end{deluxetable}
}

{
\begin{deluxetable}{lcccccccc}
\tabletypesize{\tiny}
\tabletypesize{\scriptsize}
\setlength{\tabcolsep}{0.05in}
\rotate
\tablecaption{Photometry\tablenotemark{a} of Multiple System
\label{tbl-ms}}
\tablewidth{0pt}
\tiny 
\tighttable
\tablehead{
\colhead{Name} &
\colhead{RA} &
\colhead{Dec} &
\colhead{F125W} &
\colhead{F160W-F125W} &
\colhead{F140W-F125W} &
\colhead{F105W-F125W} &
\colhead{F814W-F125W} &
\colhead{$\mu$\tablenotemark{b}} \\
\colhead{} &
\colhead{(J2000)} &
\colhead{(J2000)} &
\colhead{} &
\colhead{} &
\colhead{} &
\colhead{} &
\colhead{} &
\colhead{} 
}
\startdata
 ZD7A1  &   3.592410 & -30.409897 & $  27.33 \pm  0.08 $& $ -0.08 \pm 0.10 $& $ -0.05 \pm  0.10 $& $ 0.38 \pm  0.11 $& $> 4.0 $& $5.7 ^{+7.5} _{-2.7}$ \\ 
 ZD7A2  &  3.592160 & -30.409925 & $ 28.24 \pm 0.12 $& $ -0.06 \pm 0.16 $& $ 0.09 \pm 0.18 $& $  0.36 \pm 0.18 $& $> 3.0 $& $6.4 ^{+13.2} _{-3.4}$ \\
 ZD7B  & 3.588430  & -30.410340 & $ 27.7 \pm  0.1 $& $ -0.3 \pm  0.1 $& $  0.0 \pm  0.1 $& $ 0.6 \pm 0.3 $& $> 2.0 $& $32.5 ^{+31} _{-32}$ \\   
 ZD7C  &   3.600940 & -30.400824 & $  28.58 \pm  0.15 $& $ 0.25 \pm  0.22 $& $ -0.21 \pm  0.22 $& $ 0.81 \pm 0.28 $& $> 2.0 $& $2.8 ^{+1.5} _{-0.1}$ \\ 
\enddata
\tablenotetext{a}{See notes in Table \ref{tbl-z7}}
\end{deluxetable}
}

\begin{thebibliography}{}
\bibitem[Atek et al. (2014)]{atek} Atek, H., Richard, J., Kneib, J.-P. et al. 2014, \apj, 786, 60 (A14) \bibitem[Beckwith  et al.(2006)]{beckwith} Beckwith, S. V. W., Stiavelli, M., Koekemoer, A. M. et al. 2006, \aj, 132, 1729
\bibitem[Ben\'{\i}tez(2000)]{bpz} Ben\'{\i}tez, N, 2000, \apj, 536, 571
\bibitem[Bertin \& Arnouts(1996)]{bertin} Bertin, E., \& Arnouts, S.\ 1996, \aaps, 117, 393
\bibitem[Blakeslee et al.(2003)]{blakeslee} Blakeslee, J.~P., Anderson, K.~R., Meurer, G.~R.,
Ben{\'{\i}}tez, N., \& Magee, D.\ 2003, in Astronomical Data Analysis Software and Systems XII,
ASP Conference Series, Vol. 295, eds. H. E. Payne, R. I. Jedrzejewski, \& R. N. Hook, (San Francisco: ASP) 257
\bibitem[Bouwens et al.(2012b)]{bouwens2} Bouwens, R.~J., Bradley, L. D., Zitrin, A. et al.\ 2012$b$, arXiv 1211.2230
\bibitem[Bouwens et al.(2011)]{bouwens} Bouwens, R.~J., Illingworth, G.~D., Labbe, I. et al.\ 2011, Nature, 469, 504
\bibitem[Bouwens et al.(2010)]{bouwens3} Bouwens, R.~J., Illingworth, G.~D., Oesch, P. A. et al.\ 2010, \apj, 708, L69
\bibitem[Bouwens et al.(2012a)]{bouwens1} \underline{\hskip 7 em} 2012$a$, \apj, 737, 90 
\bibitem[Bouwens et al.(2014)]{bouwens4} \underline{\hskip 7 em} 2014, arXiv 1403.4295 \bibitem[Bradley et al.(2008)]{a1689} Bradley, L. D., Bouwens, R. J., Ford, H. C. et al.\ 2008, \apj, 678, 647 
\bibitem[Bradley et al.(2012)]{a1703} Bradley, L. D., Bouwens, R. J., Zitrin, A. et al.\ 2012, \apj, 747, 3 
\bibitem[Bradley et al.(2014)]{bradley} Bradley, L. D., Zitrin, A., Coe, D. et al.\ 2014, \apj, in press, arXiv 1308.1692
\bibitem[Chabrier (2003)]{chabrier} Chabrier, G. 2003, \pasp, 115, 763
\bibitem[Charlot \& Fall (2000)]{charlot00} Charlot, S. \& Fall, S. M. 2000, \apj, 539, 718
\bibitem[Coe et al.(2006)]{coe06} Coe, D., Ben\'itez, N., S\'anchez, 
S. F., Jee, M., Bouwens, R. \& Ford, H. 2006, \apjs, 132, 926
\bibitem[Coe et al.(2013)]{coe} Coe, D., Zitrin, A., Carrasco, M. et al. 2013, \apj, 762, 32
\bibitem[Coe et~al.(2014)]{coe14} Coe, D., Bradley, L. \& Zitrin, A., 2014, arXiv 1405.0011 \bibitem[Conroy \& Gunn (2010)]{conroy10} Conray, C. \& Gunn, J. E. 2010, \apj, 712, 833
\bibitem[Conroy, Gunn \& White (2009)]{conroy09} Conray, C., Gunn, J. E. \&  White, M. 2009, \apj, 699, 486
\bibitem[Cypriano et al.(2004)]{cypriano} Cypriano, E. S., Sodr\'e, L. Jr., Kneib, J.-P. \& Campusano, L. E. 2008, \apj, 613, 95
\bibitem[Ellis et al.(2013)]{ellis} Ellis, R.~S., McLure, R.~J., Dunlop, J.~S. et al.\ 2013, \apjl, 763, L7
\bibitem[Fazio et al. (2004)]{irac}  Fazio, G. G., Hora, J. L., Allen, L. E. et al., 2004, \apjs, 154, 10
\bibitem[Ferguson \& McGaugh (1995)]{ferguson} Ferguson, H. C. \& McGaugh, S. S. 1995, \apj, 440, 470 
\bibitem[Fioc \& Rocca-Volmerange (1997)]{pegase} Fioc, M., \& Rocca-Volmerange, B. 1997, \aap, 326, 950
\bibitem[Ford et al. (1998)]{acs} Ford, H. C. \& the ACS Science Team 1998, in Space Telescopes and Instruments V, 
SPIE 3356, eds. P. Y. Bely \& J. B. Breckinridge, 234
\bibitem[Hack, Dencheva \& Fruchter (2013)]{astrodrz} Hack, W. J., Dencheva, N., Fruchter, A. S., 2013 in Astronomical Data Analysis 
Software and Systems XXII, ASP conf. Ser. 475, (San Francisco: ASP) ed. D. Freidel, 49
\bibitem[Illingworth et al.(2013)]{garth} Illingworth, G.~D., Magee, D., Oesch, P. A. et al. 2013, \apjs, 209, 6
\bibitem[Jouvel et al.(2014)]{jouvel} Jouvel, S., H{\o}st, O., Lehav, O., et al. 2014, \aap, 562, A86
\bibitem[Kimble et al. (2008)]{wfc3} Kimble, R. A., MacKenty, J. W., O'Connell, R. W., \& Townsend, J. A. 2008,
  in Space Telescopes and Instrumentation 2008: Optical, Infrared, and Millimeter, SPIE 7010,   
  eds. J. M. Oschmann, Jr.,  M. W. M. de Graauw, \& H. A. MacEwen, 70101E
\bibitem[Koekemoer et al.(2013)]{koekemoer} Koekemoer, A.~M., Ellis, R. S., McLure, R. J. et al.\ 2013, \apjs, 209, 3
\bibitem[Labb{\'e} et al.(2006)]{labbe06} Labb{\'e}, I., Bouwens, R., Illingworth, G.~D., \& Franx, M.\ 2006, \apjl, 649, L67 
\bibitem[Labb\'e et al. (2010)]{labbe10}Labb\'e, I, Gonz\'alez, V., Bouwens, R. J. et al. 2010, \apj, 716, 103
\bibitem[Lam et al. (2014)]{lam}Lam, D., Broadhurst, T., Diego, J. M. et al. 2014, arXiv 1406.6586
\bibitem[Laporte et al.(2014)]{laporte} Laporte, N., Streblyanska, A., Clement, B. et al. 2014, \aap, 562, 8L
\bibitem[Makovoz \& Khan (2005)]{mopex} Makovoz, D. \& Khan, I. 2005, in Astronomical Data Analysis
Software and Systems VI, ASP Conf. Ser. 132, eds. P. L. Shopbell, M. C. Britton, \& R. Ebert (San Francisco: ASP), 81
\bibitem[Merten et al. (2011)]{merten} Merten, J., Coe, D., Dupke, R. et al. 2011, \mnras, 417, 333
\bibitem[Moustakas et al. (2013)]{ised} Moustakas, J., Coil, A., Aird, J. et al. 2013, \apj, 767, 50
\bibitem[Moster et al. (2011)]{moster} Moster, B. P., Somerville, R. S., Newman, J. A. \&
Hans-Walter, R. 2011, \apj, 731, 113
\bibitem[Navarro, Frenk \& White (1996)]{nfw} Navarro, J. F., Frenk, C. S. \& White, S. D. M. 1996, \apj, 462, 563
\bibitem[Oesch et al. (2010)]{oesch10} Oesch, P. A., Bouwens, R.~J., Illingworth, G.~D. et al. 2010, \apj, 709, L16
\bibitem[Oesch et al.(2013)]{oesch} \underline{\hskip 7em} 2013, \apj, 773, 75
\bibitem[Okabe, Okura \& Futamase (2010)]{okabe10} Okabe, N., Okura, Y. \&  Futamase, T. 2010, \apj, 713, 291
\bibitem[Okabe et al. (2010)]{okabe10b} Okabe, N., Takada, M., Umetsu, K. et al. 2010, \pasj, 62, 811
\bibitem[Okabe \& Umetsu (2008)]{okabe08} Okabe, N. \& Umetsu, K. 2008, \pasj, 60, 345
\bibitem[Overzier et al.(2009)]{overzier} Overzier, R. A., Shu, X., Zheng, W. et al. 2009, 704, 548 
\bibitem[Papovich et al. (2011)]{papovich11} Papovich, C., Finkelstain, S. L., Ferguson, H. C., Lotz, J. M., \& Giavalisco, M. 2011, \mnras, 412, 1123 
\bibitem[Peng et al. (2010)]{galfit}Peng, C.~Y., Ho, L. C., Impey, C. D. \& Rix, H.-W. 2010, \aj, 139,  2097
\bibitem[Postman et al.(2012)]{postman} Postman, M., Coe, D., Ben{\'{\i}}tez, N. et al.\ 2012, \apjs, 199, 25
\bibitem[Redlich et al. (2012)]{redlich} Redlich, M., Bartemann, M., Waizmann, J.-C. \& Fedeli, C. 2012, \aap, 547, 66
\bibitem[Richard et al. (2011)]{a383} Richard, J., Kneib, J.-P., Ebeling, H. et al. 2011, \mnras, 414, L31
\bibitem[Sanchez-Bl\'azquez et al. (2006)]{sanchez-blazquez06} Sanchez-Bl\'azquez, P., Peletier, R. F., Jim\'enez-Vicente, J. et al. 2006, \mnras, 371, 703
\bibitem[Torri et al. (2004)]{torri} Torri, E., Meneghetti, M., Bartelmann, M. et al. 2004, \mnras, 349, 476
\bibitem[Trenti et al.(2012)]{trenti} Trenti, M., Bradley, L. D., Stiavelli, M. et al.\ 2012, \apj, 746, 55
\bibitem[Trenti \& Stiavelli (2008)]{cv} Trenti, M. \& Stiavelli, M.  2008, \apj, 676, 767
\bibitem[Wuyts et al.(2008)]{fireworks} Wuyts, S., Labb\'e, I., Schreiber, N. et al.\ 2008, \apjs, 682, 985
\bibitem[Zheng et al.(2012)]{aplus} Zheng, W., Bradley, L. D., Saraff, A. et al.\ 2012, in
Seventh Conference on Astronomical Data Analysis, http://ada7.cosmostat.org/proceedings.php, 17
\bibitem[Zheng et al.(2012)]{zheng} Zheng, W., Postman, M., Zitrin, A. et al.\ 2012, Nature, 489, 406
\bibitem[Zitrin et al. (2009)]{zitrin} Zitrin, A., Broadhurst, T., Umetsu, K. et al. 2009, \mnras, 396, 1895
\bibitem[Zitrin et al.(2013a)]{elgordo} Zitrin, A., Menanteau, F., Hughs, J. P. et al. 2013$a$, \apj, 770, L15
\bibitem[Zitrin et al.(2013b)]{zitrin13} Zitrin, A., Meneghetti, M., Umetsu, K. et al. 2013$b$, \apj, 762, 30
\bibitem[Zitrin et al.(2014)]{zitrin14} Zitrin, A., Zheng, W., Broadhurst, T. et al. 2014,  arXiv 1407.3769
\end{thebibliography}
\end{document}